\documentclass[12pt]{article}

\newcommand{\be}{\begin{equation}}
\newcommand{\ee}{\end{equation}}
\newcommand{\bea}{\begin{eqnarray}}
\newcommand{\eea}{\end{eqnarray}}

\newcommand{\bit}{\begin{itemize}}
\newcommand{\eit}{\end{itemize}}
\newcommand{\no}{\noindent}

\begin{document}
{\sf \title{$G_2$ holonomy metrics and wrapped D6-branes}
\author{Ansar Fayyazuddin\footnote{email: Ansar\_ Fayyazuddin@baruch.cuny.edu} $^1$, Tasneem Zehra Husain\footnote{email: tasneem@physics.harvard.edu} $^2$}
\maketitle
\begin{center}
\vspace{-1cm}
{\it $^1$ Department of Natural Sciences, Baruch College, \\
City University of New York, New York, NY\\ 
$^2$ Jefferson Physical Laboratory, Harvard University, \\
Cambridge, MA 02138}
\end{center}

\begin{abstract}
We determine torsion class constraints for the supergravity background produced by D6-branes wrapping special Lagrangian cycles in a Calabi-Yau 3-fold.  We employ a recently introduced method which involves probing the putative background by all possible supersymmetric brane configurations.  We then lift this background to 11-dimensions to a product of 4-d Minkowski space and a 7-fold of $G_2$-holonomy.   The latter is a particular U(1) bundle over an almost complex manifold of SU(3) structure with specific torsion class constraints.  We construct the closed 3- and 4-forms which calibrate the 3- and 4-cycles in the $G_2$-holonomy manifold.   
\end{abstract}

\vspace{-16cm}
\begin{flushright}
HUTP-06/A0034 \\
BCCUNY-HEP /06-04 \\
hep-th/0608163
\end{flushright}

\thispagestyle{empty}

\newpage

\tableofcontents

\section{Introduction}

Supersymmetric solutions of supergravity theories have long played a distinguished role in string theory because they provide us with settings in which string theory can be studied in a controlled fashion.  In backgrounds with zero flux the classification of supersymmetric backgrounds is completely known.  For instance in 11-d supergravity, ${\bf R}^{1,10 - n} \times {\cal M}_{n}$ is a supersymmetric solution
 only if the compactification manifold ${\cal M}$ has special holonomy. There is a complete classification of such groups due to Berger and, given the dimension $n$ of ${\cal M}$, we can say immediately what sort of a manifold it has to be; if n = 2m, it must be Calabi-Yau (with SU(m) holonomy),  if n = 4m, it must be Hyper-K{\"a}hler (with Sp(m) holonomy), if n = 7, the manifold must have $G_2$ holonomy and lastly, if n = 8, the compactification manifold has reduced holonomy group Spin(7). Charged branes source a field strength for the gauge potential they couple to. Hence, supergravity backgrounds which describe branes {\it do} contain flux and the neat classification scheme described above breaks down. It has thus  been the object of much research in recent years, to come up with an analogous exhaustive list of possible manifolds in more realistic supergravity backgrounds where the flux is turned on.

%Even in the absence of a complete classification of these more general manifolds, we do have certain tools at hand. In particular,  we have a means of categorising the backgrounds when we find them. The basic idea is to use the same tensors that characterize special holonomy manifolds, but to relax the differential conditions they usually satisfy. The manifolds described by these modified forms is said to admit  a G-structure, the  'intrinsic torsion' of which gives a measure of how far the manifold deviates from having special holonomy \cite{gst}. 
The classification scheme into which we fit the system we study\footnote{For recent reviews of what is known about the classification of flux compactifications see \cite{grana,
gst}}  is that of intrinsic torsion classes of SU(3)-structures \cite{CS}.  
 The subject of our study has two alternate descriptions.  The first is in Type IIA string theory, where it takes the form of D6-branes
 wrapping a Special Lagrangian 3-cycle in a Calabi-Yau 3-fold. The second description is from the 11-dimensional point of view in
which this system appears to be pure geometry -  it is simply the product of 4-d Minkowski space and a $G_2$-holonomy manifold in
M-theory (see \cite{g} for a review of wrapped D6-branes and their lifts to special holonomy manifolds).  This problem is
 discussed in \cite{d61,d62,d63} from a different point of view: the SU(3) structure is deduced from requiring $G_2$-holonomy in
 M-theory and expressing these constraints in terms of SU(3) structures.

We will apply a recently introduced technique \cite{fh} to classify the backgrounds created by wrapped branes.  As we illustrate in the present work, this method provides us with an efficient way of finding the torsion class constraints on the geometry.  The method has the additional advantage of providing a physical meaning to these constraints - an insight which is often hard to glean when using more traditional methods involving supergravity Killing spinor equations.  

\section{Classification scheme using SU(3) structures}
In this section we briefly summarize the ideas contained in \cite{CS} that we will use to classify the supergravity solutions of
 wrapped D6-branes.  These ideas go by the name of G-structures and have been the subject of much research recently 
(see \cite{grana, gst} for a fairly recent review and references).  G in the case at hand is SU(3).  We will motivate why
we  consider SU(3) structures and then outline the classification scheme of \cite{CS}.

As explained in \cite{SpelLs}, when we consider branes wrapped on Special Lagrangian cycles (or, for that matter,  holomorphic cycles) of Calabi-Yau manifolds it is natural to assume that the Calabi-Yau manifold is replaced by an almost complex manifold.  This conclusion follows from insisting that the supersymmetry preservation conditions in the probe approximation for the wrapped branes continue to have a meaning in the full supergravity solution \cite{SpelLs}.  The existence of an almost complex structure means that forms can be decomposed into sums of (p,q) forms.  A (p,q)-form can be written as a sum 
\be
T = T_{m_1.... m_p\bar{n_1}...\bar{n_q}}e^{m_1}\wedge...\wedge e^{m_p}\wedge e^{\bar{n_1}}\wedge...\wedge e^{\bar{n_q}}
\ee
where $\{e^m\}, m=1,...,d$ is a basis of (1,0)-forms and $\{e^{\bar m}\}= \{(e^m)^*\}$ are their complex conjugates and are (0,1)-forms according to the almost complex structure.  The almost complex manifold has real dimension $2d$.  We pick the basis  $\{e^m\}$ so that they provide a frame for the manifold.  The almost complex structure allows for U(d) transformations which rotate these (1,0) forms into each other while preserving the metric.  

In the case at hand $d=3$.  There are three SU(3) invariant tensors one can form out of our basis of (1,0) forms:
\bea
g_{IJ} &=& \eta_{m\bar n}(e^m_Ie^{\bar n}_J+e^m_Je^{\bar n}_I) \nonumber \\
J&=& \frac{i}{2}(e^1\wedge e^{\bar 1} + e^2\wedge e^{\bar 2} + e^3\wedge e^{\bar 3} ) \\
\Omega &=& e^1\wedge e^2\wedge e^{3} \nonumber 
\eea
Here $g$ is the metric, $J$ is a (1,1) form and $\Omega$ is a (3,0)-form.
The last of these, $\Omega$, is invariant only under SU(3) but not under U(3).  As we will discover, $\Omega$ appears in physical quantities thus implying that the U(3) structure implied by the existence of the almost complex structure is further reduced to an SU(3) structure.

There is a classification of SU(3) structure manifolds in terms of their so-called intrinsic torsion $\tau$\cite{CS}.  The intrinsic torsion
$\tau$ has five independent components \cite{CS}
\begin{equation}
\tau \in {\cal W}_1 \oplus {\cal W}_2 \oplus {\cal W}_3 \oplus {\cal 
W}_4 \oplus {\cal W}_5  \,,
\end{equation}
${\cal W}_{i}$ label the torsion classes and, as explained in \cite{CS}, are given in terms of the exterior derivatives of $J$ and $\Omega$:
\begin{equation}
\begin{array}{rclcrcl}
{\cal W}_1 & \leftrightarrow & \left[ dJ\right]^{(3,0)}\,,  & \quad &
{\cal W}_2 & \leftrightarrow & \left[ d\Omega\right]_0^{(2,2)}\,,  \\
{\cal W}_3 & \leftrightarrow & \left[ dJ\right]_0^{(2,1)}\,,  &\quad &
{\cal W}_4 & \leftrightarrow & J \wedge dJ\,,  \\
{\cal W}_5 & \leftrightarrow & \left[ d\Omega\right]^{(3,1)}\,. & &&&
\end{array}
\label{classes}
\end{equation}
where the subscript $0$ denotes primitive forms, e.g. if $\beta \in \Lambda_{0}^{(2,2)}$ then $J \wedge \beta = 
0$, and if $\gamma \in \Lambda_{0}^{(2,1)}$ then $J \wedge \gamma =0$.

\section{How Brane Probes Help Characterize a Background}

In purely geometric backgrounds, minimal volume cycles are stable, or calibrated. Branes wrapped on such cycles have
minimal  energy/mass and therefore have nothing to decay into. Using this observation we extend the concept of calibrations to
 more general  backgrounds, defining calibrated forms to be those  which give us lower bounds on the mass, even in backgrounds
 where fields other than the metric are turned on.  Using D-brane actions to define calibrating forms has also been considered independently in \cite{martucci}. 

Since we are interested in characterizing the supergravity background produced by a wrapped brane, our strategy  is as follows. We start with an ansatz for the metric for the wrapped brane configuration.  Given this putative supergravity background  we probe it with all possible branes which preserve supersymmetry.  Such branes, being BPS, are not only stable but also static.  We compute their mass (or tension) and associate it with a calibrating form integrated over the cycle wrapped by the probe \cite{fh}. In what follows, we use this method to identify the 
calibrating forms in the supergravity background generated by D6-branes wrapping a Special Lagrangian 3-cycle. The properties of these calibrating forms give the torsion class constraints \cite{fh} and thereby allows us to characterize the background.  

We will identify the mass of the probe brane through the action of the probe.  Since the brane is static, the mass is simply given by the Lagrangian density evaluated in the supergravity background produced by the wrapped brane and then integrated over the spatial part of the probe's worldvolume.  It is this mass which we will use to find calibrating forms.  We now briefly describe the actions for the different kinds of branes we will be using as probes.  
 
The action of a D-brane in the so-called "string frame", when no further worldvolume fields are present, is given by the volume form on the worldvolume multiplied by an overall factor of the dilaton and integrated over the worldvolume of the brane\footnote{We are assuming that the pullback of the NS-NS B-field onto the worldvolume also vanishes.}:
\be
S = T_p\int e^{-\phi}\sqrt{\det h} \;d\sigma^0\wedge .... \wedge d\sigma^p
\ee
Here $\sigma^i$ are worldvolume coordinates, $\phi$ is the dilaton, $T_p$ is the tension of the Dp-brane in 10-d and $h$ is the pullback of the spacetime metric onto the D-brane.
 
The situation is slightly more complicated  for NS5-branes in type IIA string theory because there are now additional terms in the action . One way of finding these extra terms is by viewing the NS5-brane as an M5-brane in M-theory on a circle bundle over the type IIA string-frame geometry:
\begin{equation}
ds_{11}^2 = e^{-2\phi /3}ds^2_{10} + e^{4\phi /3}(dy + A_i dx^i)^2 \label{g11}
\end{equation}
Here the coordinate $y$ is along the circle and $A_i$ is the R-R 1-form in the 10-d space-time of the type IIA background.  Since the M5-brane descends to a NS5-brane only if it is transverse to the circle, we will assume that to be the case in what follows. For an M5-brane in a purely geometric background (i.e. with a vanishing 3-form), the worldvolume metric is given by the pullback of the spacetime metric (\ref{g11}):
\begin{equation}
e^{-2\phi /3}(h_{ab}+  e^{2\phi}a_aa_b)=\partial_aX^i\partial_bX^j(e^{-2\phi /3}g_{ij} + e^{4\phi/3}A_iA_j)
\end{equation}
where $h$ is the pullback of the 10-d string frame metric $g$ and $a_b$ is the pullback of $A_i$ on to the M5-brane. The action is then given by the volume form, integrated over the entire NS5-brane. Using the fact that the volume form on the M5-brane is
\be
\sqrt{ \det [e^{-2\phi /3}(h_{ab} +  e^{2\phi}a_aa_b)]} \; d\sigma^0\wedge ....\wedge d\sigma^{5} \nonumber 
\ee
and the identity 
\begin{equation}
\det(h_{ab}+  e^{2\phi}a_aa_b) = (1+e^{2\phi}a_aa_b h^{ab})\det{h} 
\end{equation}
the action can be expressed quite simply, as:
\begin{equation}
S_{NS5} = T_5\int e^{-2\phi}\sqrt{(1+e^{2\phi}a_aa_b h^{ab})\det{h} }d\sigma^0\wedge ....\wedge d\sigma^{5}. \label{ns5}
\end{equation}

Given these actions, we now have a concrete method of computing the mass of any D-brane or NS5-brane which can be introduced as a supersymmetric probe into our background. In each case, the mass will be given by the integral of certain forms. By requiring these forms to be calibrations, we will arrive at a set of contraints which classify the background. 

\section{Probing a D6-brane on a SpelL3-cycle}

The background we want to probe is one created by D6-branes wrapping a SpelL 3-cycle in a Calabi-Yau 3-fold.  We introduce the following ansatz for the space-time metric : 
\be
ds^2_{10} = H^2 \eta_{\mu \nu} dx^{\mu} dx^{\nu} + g_{IJ} dy^I dy^J \label{g10}
\ee
Here  $x^{\mu}, {\mu} = 0123$ are the coordinates on the D6-brane transverse to the Calabi-Yau, and $y^I, I = 4, \dots 9$ are coordinates on the 6-dimensional manifold we call ${\cal M}$ that approaches the underlying Calabi-Yau manifold as we "turn off" the D6-branes. Since the 3+1 dimensional space-time of the D6-branes transverse to the Calabi-Yau is Poincare invariant, the warp factor $H = H(y)$ is a function only of the coordinates on the six-manifold ${\cal M}$ with metric $g_{IJ}$.  

In addition to the above ansatz we make the important assumption that ${\cal M}$ has an almost complex
structure
 defined on it \cite{SpelLs}.  This means that there is a U(3) structure which may be reduced to an SU(3) structure. The almost
complex  structure on ${\cal M}$ then allows us to define $J$ and $\Omega$.  In the background
produced by the D6-branes we can introduce supersymmetric probes.  We will wrap them on supersymmetric cycles of ${\cal M}$.  These configurations will have an interpretation as a brane wrapped on a supersymmetric cycle of the underlying Calabi-Yau.  We will require that the configuration is supersymmetric, which for D-branes means that:
\be
\epsilon_L = \frac{1}{p!}\Gamma^{A_1...A_{p+1}}\epsilon_{A_1...A_{p+1}}\epsilon_R
\ee
Here $\epsilon_{A_1...A_{p+1}}$ is the volume form on the worldvolume of the probe brane.  Using the almost complex structure
 it is straightforward to show that supersymmetry requires that the volume of the wrapped cycles are given by the pullbacks of either
 ${\sf Re} \; [{e^{i\alpha}\Omega}], J$ or $\frac{1}{2} J\wedge J$ depending on whether the brane is wrapping a cycle
corresponding
to a Special Lagrangian cycle, holomorphic 2-cycle, or holomorphic 4-cycle, respectively, in the underlying Calabi-Yau.

The D6-branes we are interested in wrap a special Lagrangian 3-cycle calibrated by ${\sf Re}\; \Omega$. 
This can be schematically represented as:
\be
\begin{array}[h]{|c|cccc|cccccc|}
\hline
   \; & 0 & 1 & 2 & 3 & 
              4 & 5 & 6 & 7 & 8 & 9 \\
  \hline
  {\bf D6} & \otimes & \otimes & \otimes & \otimes & \otimes & \otimes  
               & \otimes  &  &  & \\
  \hline
\end{array}
\ee
We can now introduce probe branes into this background and try to study them as objects in the worldvolume theory on the non-compact part of the D6-branes.  We will find that, just as in \cite{fh}, we get constraints on $J$ and $\Omega$ giving us important information about the manifold ${\cal M}$.  

\subsection{D-brane Probes}

\no
The simplest BPS probe we can introduce into this background, is a D2-brane completely transverse to ${\cal M}$
\be
\begin{array}[h]{|c|cccc|cccccc|}
  \hline
   \; & 0 & 1 & 2 & 3 & 
              4 & 5 & 6 & 7 & 8 & 9 \\
  \hline
  {\bf D6} & \otimes & \otimes & \otimes & \otimes & \otimes & \otimes  
               &  \otimes &  &  &  \\
{\it D2} & \times & \times & \times & &  & & & &  &  \\
  \hline
\end{array}
\ee
This D2-brane appears as a flat 2-brane in the 4-d worldvolume of the D6-brane transverse to ${\cal M}$. The action of this probe is given by 
\be
S = T_2\int H^3 \; e^{ -\phi} dt\wedge dx^1\wedge dx^2
\ee
and its tension  is:
\be
T=T_2H^3 \; e^{ -\phi}.
\ee
As argued in \cite{fh} the tension for a supersymmetric probe brane is given by a calibrating form integrated over the cycle the brane is wrapping.  In this case the D2-brane is wrapping a 0-cycle and therefore the tension is given by a calibrated form only if 
\be
d_6 [e^{ -\phi} H^3] = 0
\ee
From this condition, we can read off 
\be
e^{-\phi} = H^{-3} \label{dilaton}
\ee
where we have absorbed the asymptotic value of the dilaton in $T_2$.  We will use this identity in what follows. As an additional check, recall that the lift to M-theory of our wrapped D6-brane configuration is pure geometry (see for instance \cite{g} for a comprehensive review of lifts of D6-branes to geometry).  The 11-d geometry is a product R$^{3,1}\times\cal{N}$, where $\cal{N}$ is a G$_2$-holonomy manifold.  Using this product form, and equations  (\ref{g11}) and (\ref{g10}) we can read off the relation $e^{2\phi/3}=H^2$.  This agrees with our identification using the brane probe method.

Next, we introduce a D4-brane probe which wraps a holomorphic cycle in the underlying Calabi-Yau.  In order for this probe to be BPS it must be oriented so that the total number of ND directions of the system is 0 mod 4. This is accomplished by, for instance, the following configuration:

\be
\begin{array}[h]{|c|cccc|cccccc|}
\hline
  \; & 0 & 1 & 2 & 3 & 
              4 & 5 & 6 & 7 & 8 & 9 \\
  \hline
  {\bf D6} & \otimes & \otimes & \otimes & \otimes & \otimes & \otimes  
               &  \otimes &  &  &  \\
{\it D4} & \times & \times & \times & & \times &  &  & \times & & \\
\hline
\end{array}
\ee
Supersymmetry requires that the volume form of a minimal 2-cycle in ${\cal M}$ is given by the (1,1)-form $J$. 
The action of the D4 probe is given by the integral
\be
S = T_4\int H^{3} \; e^{ -\phi} \;  dt\wedge dx^1\wedge dx^2\wedge J
\ee
The factor of $H^3$ comes from the determinant of the metric in the $0,1,2$ directions.  This D4-brane appears as a 2-brane in the flat part of the D6-brane transverse to the original Calabi-Yau.  The tension of this 2-brane inside the 4-d worldvolume theory of the D6-brane is given by:
\be
T=T_4\int_{\Sigma_2}H^{3} \; e^{ -\phi} J = T_4\int_{\Sigma_2}J
\ee
where the second equality follows from (\ref{dilaton}).  Using the arguments outlined in \cite{fh} we conclude:
\be
d_6  J = 0 \label{j},
\ee
that is, the (1,1) form $J$ is closed.

We now turn to a third possibility: a D4-brane wrapping a SpelL3-cycle. This  configuration perserves supersymmetry if the probe is oriented such that it wraps a cycle calibrated by Im $\Omega$ (recall that the D6-brane wraps a SpelL3-cycle, calibrated by Re $\Omega$).  Diagramatically, we can represent
the set-up as follows
\be
\begin{array}[h]{|c|cccc|cccccc|}
  \hline
   \; & 0 & 1 & 2 & 3 & 
              4 & 5 & 6 & 7 & 8 & 9 \\
  \hline
  {\bf D6} & \otimes & \otimes & \otimes & \otimes & \otimes & \otimes  
               &  \otimes &  &  &  \\
{\it D4} & \times & \times & & &  & & & \times & \times & \times \\
  \hline
\end{array}
\ee
The volume form on the SpelL3-cycle is the pullback of Im$\Omega$.  The action of the probe is then:
\be
S = T_4\int_{\Sigma_3\times R^{1,1}} H^2 \; e^{ -\phi} \; {\sf Im} \Omega\wedge dt\wedge dx^1
\ee
where $\Sigma_3$ is the SpelL 3-cycle the D4-brane wraps.  In the 4-dimensional worldvolume theory of the D6-brane transverse to the Calabi-Yau, this D4-brane appears to be a string with tension of this string :
\be
T = T_4\int_{\Sigma_3} H^2 \; e^{ -\phi} \; {\sf Im} \Omega = T_4\int_{\Sigma_3} H^{-1} \; {\sf Im} \Omega
\ee
where the second equality follows from (\ref{dilaton}).  Following the logic which should by now be familiar, we reason that there is a calibrating form for this tension, given by the quantity integrated over in the above expression.  Moreover, this calibration is closed:
\be
d_6 [H^{-1} \; Im \Omega] = 0, \label{imomega}
\ee
This provides us with a new torsion class constraint. We find that:
\be
d(H^{-1}\Omega) \in \Lambda^{(2,2)}.
\ee 
In general $d(H^{-1}\Omega)$ can also have a component in $\Lambda^{(3,1)}$. There is no such piece in our present case.  Moreover, the condition (\ref{imomega}) tells us that $d(H^{-1}\Omega)$ is real.  We will return to this condition later when we summarize the torsion class constraints derived in this section.

Note that a D4-brane probe with worldvolume 01457 would also describe a D4-brane wrapping a SpelL3-cycle in ${\cal M}$, calibrated by Im $\Omega$. As one would expect, the D4-probe, even in this new orientation would lead to the same calibration obtained above. Other supersymmetric D-brane probes can also be introduced, but these do not lead to any new constraints. Consider, for example, a D6-brane wrapping a 4-cycle in ${\cal M}$. 
\be
\begin{array}[h]{|c|cccc|cccccc|}
 \hline
   \; & 0 & 1 & 2 & 3 & 
              4 & 5 & 6 & 7 & 8 & 9 \\
  \hline
  {\bf D6} & \otimes & \otimes & \otimes & \otimes & \otimes & \otimes  
               &  \otimes &  &  &  \\
{\it D6} & \times & \times & \times & & \times & \times & & \times & \times &  \\
  \hline
\end{array}
\ee
This BPS probe is a 2-brane in the worldvolume theory of the D6-brane transverse to the Calabi-Yau.  It has an action:
\be
S = T_6\int H^3 \; e^{ -\phi} \; J \wedge J\wedge dt\wedge dx^1\wedge dx^2
\ee
from which we can read off the tension of the 2-brane:
\be
T= T_6\int_{\Sigma_4} H^3 \; e^{ -\phi} \; J \wedge J = T_6\int_{\Sigma_4} J \wedge J
\ee
which leads to the constraint
\be
d_6 [J \wedge J] = 0.
\ee
This is not a new constraint, it follows from the condition (\ref{j}) found above. 

Having exhausted the calibrations we can obtain using D-branes, we now turn to the somewhat more complicated NS-probes. 

\subsection{NS-Brane Probes}

For the case at hand, NS5-branes can only be introduced in two ways such that supersymmetry is preserved. In the first scenario, a NS5-brane wraps a SpelL 3-cycle, and in the second, it wraps a holomorphic 4-cycle.  

Consider the SpelL wrapping first.  For the configuration to be supersymmetric the 3-cycle $\Sigma_3$ must be calibrated by Re $\Omega$ (the same as the background D6-brane).  Visually, we can represent this as follows:
\be
\begin{array}[h]{|c|cccc|cccccc|}
 \hline
   \; & 0 & 1 & 2 & 3 & 
              4 & 5 & 6 & 7 & 8 & 9 \\
  \hline
  {\bf D6} & \otimes & \otimes & \otimes & \otimes & \otimes & \otimes  
               &  \otimes &  &  &  \\
{\it NS5} & \times & \times & \times & & \times & \times & \times &  &  &  \\
  \hline
\end{array}
\ee
The action for our NS5-brane probe is (\ref{ns5}):
\begin{equation}
S_{NS5} = T_5\int e^{-2\phi}\sqrt{ {\sf det} h \; (1+e^{2\phi}a_aa_b h^{ab})}  
dt\wedge dx^1\wedge dx^2\wedge d\sigma^1\wedge d\sigma^2\wedge d\sigma^{3}\label{ns52}
\end{equation}
where we have split the worldvolume coordinates into $t,x^1,x^2$ transverse to ${\cal M}$ and $\sigma^1,\sigma^2,\sigma^3$ on $\Sigma_3$.  
We are interested in a static supersymmetric NS5-brane.  As one can see from the table above, the NS5-brane has two non-compact directions along the D6-brane transverse to $\cal M$.  It appears, therefore, as a 2-brane in the worldvolume theory of the D6-brane transverse to the Calabi-Yau.  The tension of this 2-brane can be read off from (\ref{ns52}) to be:
\begin{equation}
T = T_5\int e^{-2\phi}\sqrt{ {\sf det} h \; (1+e^{2\phi}a_aa_b h^{ab}) } d\sigma^1\wedge d\sigma^2\wedge d\sigma^{3}
\end{equation}
This expression can be simplified using the following observation. On a SpelL cycle the pullback of the K{\"a}hler form vanishes by definition. Although the geometry produced by the D6-brane is no longer complex it does retain the almost complex structure of the underlying Calabi-Yau and, as expected, the condition $J|_{\Sigma_3} =0$ holds .  That being the case it is straightforward, although tedious, to show that:
\begin{equation}
\det h = H^6\; |\Omega_{|\Sigma_3}|^2 =H^6[ ({\sf Re} \Omega_{|\Sigma_3})^2 + ( {\sf Im} \Omega_{|\Sigma_3})^2], 
\end{equation}
i.e. the determinant of the pullback of the 10-dimensional metric is the sum of two non-negative terms - the squares of the real and imaginary parts of the pullback of $\Omega$ onto the 3-cycle \footnote{Of course, there are an infinite set of ways of breaking up the term into two non-negative parts by multiplying $\Omega$ by a phase.}. The factor of $H^6$ comes from the three non-compact worldvolume directions of the NS5-brane, transverse to the Calabi-Yau.  The above expression enables us to re-write the tension as follows
\begin{equation}
T = T_5\int H^3 e^{-2\phi}\sqrt{ (1+e^{2\phi}a_aa_b h^{ab}) 
[({\sf Re} \Omega_{|\Sigma_3})^2 + ( {\sf Im} \Omega_{|\Sigma_3})^2)]
} d\sigma^1\wedge d\sigma^2\wedge d\sigma^{3}
\end{equation}
in a form that is suggestive of a Bogomolnyi-type bound.  
For a static configuration that cannot decay into something energetically more favorable, we have to minimize the tension. Since we have a product of two non-negative expressions under the square root, clearly the tension is minimal when both non-negative terms are minimized separately, i.e. 
\be
{\sf Re} [e^{i\alpha}\Omega_{|\Sigma_3}] =0 \;{\sf for \; some \; phase} \; \alpha \;\;\;\;\;\; {\sf and} \;\;\;\;\  a_aa_bh^{ab}=0
\ee
The later condition is only possible if $a= A_{|\Sigma_3}=0$.  As discussed earlier, a  NS5-brane probe in this background is supersymmetric only when calibrated by  Re $\Omega$; in other words, we put Im $\Omega_{|\Sigma_3}=0$. This reasoning allows us to determine a closed form.  Notice that the mass of the 2-brane (i.e. the NS5/D6-brane intersection in the non-compact directions) is determined by 
\begin{equation}
T = T_5\int_{\Sigma_3} H^3e^{-2\phi} \; {\sf Re} \Omega. \label{notquite}
\end{equation}

Because the RR 1-form $A$ vanishes on $\Sigma_3$, any piece in Re $\Omega$ which has a non-zero contraction with $A$ will not contribute to the mass.  For a minimum mass configuration, it is thus equally true that the tension can be computed by 
\begin{eqnarray}
T &=& T_5\int_{\Sigma_3} H^3e^{-2\phi} 
[{\sf Re} \; \Omega - \frac{1}{2}\frac{A^k}{A_mA^m} {\sf Re} \Omega_{ijk} \; dx^i\wedge dx^j\wedge A] \nonumber \\
&=& T_5\int_{\Sigma_3} H^3e^{-2\phi} [{\sf Re} \Omega - \frac{1}{A_mA^m}(*_6 [A \wedge {\sf Im} \; \Omega]) \wedge A] 
\label{spellcalib}
\end{eqnarray}
Does a calibrating form exist, associated with this NS5-brane?  Such a calibrating form should compute the minimum tension when integrated over any cycle which is homologically equivalent to the minimal cycle.  In general, a cycle homologous to the minimal one will not have $A_i$ vanishing on it and the integral (\ref{notquite}) over such a  cycle will not compute the minimum tension.  The expression (\ref{spellcalib}) however, gives the minimal tension for all $\Sigma_3$ in the same homology class as the minimal cycle but not necessarily minimal itself. Hence  the calibration in question is given by the closed form 
\begin{equation}
 H^3e^{-2\phi} [{\sf Re} \Omega - \frac{1}{A_mA^m}(*_6 [A \wedge {\sf Im} \; \Omega]) \wedge A]. 
\end{equation} 
Simplyifying this expression using (\ref{dilaton}) gives finally the constraint:
\be
d \{ H^{-3}[{\sf Re} \Omega - \frac{1}{A_mA^m}(*_6 [A \wedge {\sf Im} \; \Omega]) \wedge A]\}=0
\ee
We shall see that there is a much more elegant expression for the calibrating form when we lift to 11-dimensions. A similar analysis can be carried out for a NS5-brane wrapped on  a 4-cycle in ${\cal M}$:
\be
\begin{array}[h]{|c|cccc|cccccc|}
 \hline
   \; & 0 & 1 & 2 & 3 & 
              4 & 5 & 6 & 7 & 8 & 9 \\
  \hline
  {\bf D6} & \otimes & \otimes & \otimes & \otimes & \otimes & \otimes  
               &  \otimes &  &  &  \\
{\it NS5} & \times & \times & & & \times & \times & & \times & \times &  \\
  \hline
\end{array}
\ee
The action for this  configuration is given by adapting (\ref{ns5}) to the case at hand:
\bea
S_{NS5} &=& T_5\int e^{-2\phi}\sqrt{ {\sf det} h \; (1+e^{2\phi}a_aa_b h^{ab})}  
dt\wedge dx^1\wedge d\sigma^1\wedge d\sigma^2\wedge d\sigma^{3}\wedge d\sigma^4 \nonumber \\
&=& T_5\int e^{-2\phi}H^2\sqrt{ (1+e^{2\phi}a_aa_b h^{ab})}\frac{1}{2} J\wedge J\wedge 
dt\wedge dx^1 \nonumber \\
&=&T_5\int H^{-4}\sqrt{ (1+e^{2\phi}a_aa_b h^{ab})} \frac{1}{2}J\wedge J\wedge 
dt\wedge dx^1.
\eea
This NS5-brane appears as a string in the 3+1 dimensional non-compact part of the worldvolume of the D6-brane with tension:
\be
T = T_5\int_{\Sigma_4} H^{-4}\sqrt{ (1+e^{2\phi}a_aa_b h^{ab})} \frac{1}{2}J\wedge J
\ee
For the tension to be minimal the pullback of the R-R 1-form $A$ onto the supersymmetric 4-cycle, $\Sigma_4$ must vanish: $a_b =0$.  If that condition is satisfied then the tension is given by:
\be
T = \frac{1}{2}T_5\int_{\Sigma_4} H^{-4}J\wedge J.
\ee
Just as above we can construct a closed form $\phi$ by removing those terms in $J\wedge J$ that have a non-zero projection along $A$:
\be
\phi = H^{-4}J\wedge J - \frac{1}{3!}\frac{A^l}{A_mA^m}(J\wedge J)_{ijkl}dx^i\wedge dx^j\wedge dx^k\wedge A
\ee
Again, a more elegant expression for the calibrating form arises naturally in 11-d as we shall below.

\section{The supergravity solution for D6-branes: torsion classes and the R-R 1-form}
In this section we summarize our results from the previous section in terms of the torsion classes for the geometry.  We also complete the characterization of the supergravity solution by showing how to compute the R-R one-form $A$. 

The torsion classes are given in (\ref{classes}).  In the previous section we found that (\ref{j}) $J$ is closed.  In terms of torsion classes this restricts us to
\be
{\cal W}_1 = {\cal W}_3 = {\cal W}_4 = 0.
\ee
\no
In addition we found that the combination $H^{-1}{\sf Im}\Omega$ is closed (see (\ref{imomega})).  From this we can conclude that:
\bea
(d\Omega)^{(3,1)} &=& H^{-1}dH\wedge \Omega \\
(d\Omega)^{(2,2)} &=& (d\bar{\Omega})^{(2,2)} \label{Omega}
\eea
Thus ${\cal W}_5$ is fixed by the first of these equations.  The last of the above equations restricts ${\cal W}_2$ to be real.  

Having specified the torsion class of the metric $g$, we turn to determining $A$.  Here it is useful to combine our techniques with those of generalized calibrations
\cite{gencal}. The key idea underlying generalized calibrations is that for states saturating a BPS bound, the
mass and charge are identical. 
A BPS p-brane couples (electrically) to a $(p+1)$-form and the BPS relation between mass and charge
implies that this $(p+1)$-form gauge field is equal to the effective volume form (i.e. the Lagrange density on the
worldvolume evaluated in the supergravity background) on the (p+1)-dimensional worldvolume of the p-brane.  
In our
case we have a D6-brane with a (6+1)-dimensional worldvolume which couples electrically to a 7-form, the magnetic dual of
the R-R 1-form $A$.  Using the generalized calibration idea we can write down an expression for the 7-form $\tilde A_7$ by equating it with the DBI Lagrangian for the D6-brane - it's effective mass density:
\bea
{\tilde A_7} &=& T_6e^{-\phi}H^{4}dx^0\wedge dx^1\wedge dx^2\wedge dx^3\wedge {\sf Re}\Omega \nonumber \\
&=&T_6Hdx^0\wedge dx^1\wedge dx^2\wedge dx^3\wedge {\sf Re}\Omega 
\eea
where $T_6$ is the tension of the D6-brane in 10-d and we have used the fact that the volume of the Special Lagrangian
3-cycle is given by the pullback of ${\sf Re}\Omega$.  From the above expresion we can construct an 8-form field strength:
\be
{\tilde F}_8 = d{\tilde A_7} = T_6dx^0\wedge dx^1\wedge dx^2\wedge dx^3\wedge 
d(H\;{\sf Re}\Omega)
\ee
which is the Hodge dual of $F_2= dA$ that defines $A$ for us:
\bea
F_2 &=& *_{10}d{\tilde A_7} \nonumber \\ 
&=& H^{-4}*_6d_6(H{\sf Re}\Omega) \\
&\equiv& dA \label{F}
\eea
This last equation implicitly gives us a way of determining $A$ in terms of geometric data.  We can further analyze
the above expression for $F$ by noticing that it is a 2-form which can be decomposed into (2,0), (0,2) and (1,1) components using the almost complex structure.  The torsion class constraints (\ref{Omega}) simplify the expressions to: 
\bea
F^{(2,0)} &=& H^{-4}*_6(dH\wedge\Omega) = iH^{-4}\partial_IH\Omega^I_{\;JK}\frac{1}{2}dy^J\wedge dy^K\nonumber \\
F^{(0,2)} &=& H^{-4}*_6(dH\wedge\bar{\Omega})= (F^{(2,0)})^* \nonumber \\
F^{(1,1)} &=& H^{-3}*_6(d {\sf Re}\Omega)^{(2,2)}
\eea
Finally, $F$ has to satisfy the equation of motion: $dF=0$:
\be
0 = d_6(H^{-4}*_6d_6(H \; {\sf Re}\Omega)).
\ee
This completes our analysis of the supergravity solution for D6-branes wrapping Special Lagrangian Cycles in Calabi-Yau 3-folds.  We
now turn to the lift of these results to 11-dimensions.

\section{The 11-d lift of wrapped D6-branes and $G_2$-holonomy manifolds}

The 10 dimensional type IIA supergravity description of wrapped D6-branes can be lifted to 11-dimensions with the identification:
\begin{equation}
ds_{11}^2=e^{-{2\phi\over 3}}h_{ab}dx^a dx^b +e^{{4\phi\over
3}}(d\psi+A_I dx^{I})^2.
\end{equation}
where $h$ is the 10-dimesnional string frame metric (\ref{g10}), $A$ is the R-R 1-form and $\psi$ is a compact coordinate along the 11th dimension. 
Since the metric is invariant under translations in $\psi$, there is no explicit $\psi$ dependence. 
Generically, type
IIA solutions can have other fields turned on besides the dilaton and R-R 1-form.  Such solutions lift to 11-dimensional
solutions with a non-zero 3-form potential.  In our case there are no such fluxes present so we lift to a
purely geometric background.   

More explicitly with our identification of the dilaton in (\ref{dilaton}) and the string frame metric (\ref{g10}), 
the 11-dimensional space-time is:
\be
ds_{11}^{2} = \eta_{\mu\nu}dx^\mu dx^\nu + H^{-2}g_{IJ}dy^Idy^J + H^4(d\psi + A_Idy^I)^2
\ee
This metric is a product of $R^{3,1}$ and a 7-dimensional manifold which is a U(1) bundle over the almost complex
manifold $\cal{M}$:
\be
ds_7^2 = H^{-2}g_{IJ}dy^Idy^J + H^4(d\psi + A_Idy^I)^2 \label{g2}
\ee
We denote this 7-dimensional manifold ${\cal M}_7$.  Since we have a purely geometric supersymmetric solution in 11 dimensions with 4 supersymmetries, Berger's classification immediately tells us that the manifold ${\cal M}_7$ has $G_2$
holonomy\footnote{Manifolds with $G_{2}$ holonomy have previously been constructed from six-dimensional manifolds with
$SU(3)$ structure; see, for example \cite{misra}}. 

A $G_2$ holonomy manifold is, as its name implies, a seven dimensional manifold whose holonomy group is the simple group $G_2$. Such a manifold always admits a distinguished harmonic three-form
$\Lambda_3$ which satisfies $d \Lambda_3 = d * \Lambda_3 = 0$.   The forms $\Lambda_3$ and $\Lambda_4\equiv *\Lambda_3$
calibrate 3 and 4-cycles in ${\cal M}_7$ and are referred to as associative and co-associative
calibrations respectively.  We will now construct $\Lambda_3$ and $\Lambda_4$ using our 10-d analysis from the
previous sections. Before proceeding to the actual construction however, we pause briefly to discuss a point
that will be useful.  Physics is invariant under gauge transformations of the R-R 1-form $A$: $A\rightarrow A + d\lambda$.  This invariance is explicit in the type IIA context where the
 gauge field is defined through the field strength (\ref{F}).  Since $A$ appears in the expression for the metric (\ref{g2}),
one might wonder how gauge invariance is reflected in this context.  It turns out that this metric (\ref{g2}) is in
fact gauge invariant up to re-definitions of the coordinate $\psi$.  In particular, the existence of the Killiing
vector field $ \partial_\psi$ allows us to bring the metric back to its original form through the transformation:
\bea
A&\rightarrow& A + d\lambda \nonumber \\
\psi &\rightarrow& \psi - \lambda.
\eea
The lesson here is that while neither $A$ nor $d\psi$ is gauge invariant on its own, the combination $d\psi + A$ is. 
The significance of this observation will be apparent in a moment.
We turn now to the construction of the forms $\Lambda_3$ and $\Lambda_4$ which calibrate 3 and 4-cycles
in 
${\cal M}_7$. 
Three
cycles in ${\cal M}_7$ have two origins: a) as three cycles in ${\cal M}$ and as b) a circle ($\psi$) fibered over a two cycle in  ${\cal M}$ (this circle is pinched at zeroes of $H$).  To connect to the discussion in the previous sections in the type IIA context, consider wrapping an M5-brane on these two different types of cycles.  Case a) corresponds to NS5-branes wrapping Special Lagrangian cycles in the Calabi-Yau while case b) corresponds to D4-branes wrapping holomorphic 2-cycles in the Calabi-Yau.  We know how to compute the masses of each of these objects.  The tension of the non-wrapped part of the D4-brane in case b), as we discussed previously, is given by:
\be
T = T_4\int_{\Sigma_2}J = T_{M5}\int_{\Sigma_2\times S^1}(H^{-2}J)\wedge(H^2d\psi)
\ee
where we have made the usual identification:
\be
T_4 = T_{M5}\int_{S^1}d\psi.
\ee
Notice that we have paired the factors of $H$ suggestively to indicate their origins.  Now according to our analysis we should have a closed calibrating form:
\be
\Lambda' = J\wedge d\psi. 
\ee
This form is in fact closed but it is not gauge invariant.  We might consider modifying the above expression to:
\be
\Lambda'' = J\wedge (d\psi + A)
\ee
This form is gauge invariant by construction but it is no longer closed:
\be
d\Lambda'' = J\wedge F.
\ee
Using the explicit expression for $F$ given in equation (\ref{F}) and the torsion class constraints (\ref{Omega}), it is possible to show that:
\be
J\wedge F = -d(H^{-3}{\sf Re}\Omega).
\ee
and, therefore, 
\be
\Lambda_3 = J\wedge (d\psi + A) + H^{-3}{\sf Re}\Omega
\ee
is closed.  We have somewhat hastily identified $\Lambda_3$ with an expression which is formally closed, without
discussing the physical meaning of the additional term; this term however turns out to be easy to understand. In
section 4.2, we showed that supersymmetric NS5-branes wrapped on Special Lagrangian cycles were calibrated by $H^{-3}{\sf Re}\Omega$,
and that the pullback of $A$ vanished on these cycles.  From the M-theory point of view, this is precisely when the
only non-zero contribution comes from the terms we just added.  This what we labeled case a)
above.  

To construct $\Lambda_4$ we could simply take the 7-d Hodge dual of $\Lambda_3$ but we prefer instead to derive it in a
manner
similar to our derivation of $\Lambda_3$ so that its interpretation as a calibration is clear.  Consider the possible
origins of 4-cycles in ${\cal M}_7$.  There are again two possibilities: case a), where the 4-cycle is completely
contained inside of ${\cal M}$ and case b), when the 4-cycle is a circle fibered over a 3-cycle in ${\cal M}$.  Again, to
connect to our previous discussion consider an M5-brane wrapping a 4-cycle.  Case a) corresponds to a NS5-brane wrapping a
 holomorphic 4-cycle in the Calabi-Yau while case b) corresponds to a D4-brane wrapping a Special Lagrangian 3-cycle in the Calabi-Yau. 
 Let us compute the tension of the non-compact directions of the D4-brane in case b) as we did in section 4. We find
that:
\be
T = T_4\int_{\Sigma_3}H^{-1}{\sf Im}\Omega = T_{M5}\int_{\Sigma_3\times S^1}(H^{-3}{\sf Im}\Omega)\wedge(H^2d\psi)
\ee
where we have again grouped the factors of $H$ in a manner suggestive of their origins.  As before, the natural
calibrating 4-form for this class of configurations is:
\be
\Lambda''' = H^{-1}{\sf Im}\Omega\wedge d\psi.
\ee
While closed this is not gauge invariant, so proceeding as we did above we might consider:
\be
\Lambda ''''=H^{-1}{\sf Im}\Omega\wedge (d\psi + A).
\ee
Mirroring the discussion for the 3-form, we have now arrived at a gauge invariant version of the calibration which is not
closed:
\be
d\Lambda '''' = H^{-1}{\sf Im}\Omega\wedge F
\ee
where $F$ is given in (\ref{F}).  It is not difficult to show that:
\be
H^{-1}{\sf Im}\Omega\wedge F = -\frac{1}{2}dH^{-4}\wedge J\wedge J = -\frac{1}{2}d(H^{-4}J\wedge J).
\ee
where the last equality makes use of our torsion constraint $dJ=0$.  Thus the 4-form:
\be
\Lambda_4 = H^{-1}{\sf Im}\Omega\wedge (d\psi + A) + \frac{1}{2}H^{-4}J\wedge J
\ee
is both closed and gauge invariant.  The additional term introduced here is precisely what we needed to cover case b) as
discussed in section 4.2.  

We have now assembled all the elements we sought. 

\section{Conclusions}
In this paper we provide a general description of the supergravity solution of D6-branes wrapping Special Lagrangian cycles in
Calabi-Yau  manifolds.  We also explicate the relationship of these solutions to metrics on $G_2$-holonomy manifolds.  

Our analysis is based on a fundamental assumption concerning the geometry produced by wrapped branes - that although the complex structure of the underlying Calabi-Yau does not survive when a brane is wrapped on a Special Lagrangian cycle, there is an almost complex structure that remains intact.  We use methods advocated in our paper \cite{fh} (see also \cite{martucci}) to find the constraints on the almost complex geometry.  These constraints are expressed in terms of the SU(3) invariant objects $\Omega$ and $J$ which are distinguished (3,0) and (1,1) forms in the almost complex structure classification.  Our constraints allow us to put them in the context of SU(3) structures of \cite{CS}.  Beyond the constraints on the geometry we express other supergravity fields, in this case the R-R 1-form and dilaton, in terms of these objects.  To do this we couple our methods with those of generalized calibrations \cite{gencal}. 

Our methods are a generalization of \cite{fh} to string theory.  These methods are different from those usually employed in finding supersymmetric supergravity solutions.  We don't make use of Killing spinor equations in supergravity.  We instead posit a putative metric and find constraints on it by probing the background by all possible objects that have an interpretation in the flat part of the worldvolume theory of the wrapped brane.  Part of the reason for writing this paper is to advertise this method. There are several advantages to it.  The first, as we hope we have convinced readers, is of the efficiency of the method in relation to Killing spinor methods.  Furthermore, our method gives a physical meaning to the often opaque constraints one obtains on geometric structures from Killing spinor considerations.  Directly relating the constraints to the idea of a calibrated intersection makes the reason behind the constraint, if not transparent, then at least less mysterious.  

We lift the solutions for wrapped D6-branes to 11-d where we find a product of a $G_2$-holonomy geometry with 4-d flat Minkowski space.   We show how these special $G_2$-holonomy manifolds can be viewed as U(1) bundles over an almost complex 6-fold.  We show how the U(1) bundle can be expressed in terms of geometric structures on the 6-fold.  We construct explicitly calibrating 3- and 4-forms on the $G_2$-holonomy manifold.  

Our results are related to those of \cite{d61,d62,d63}.  In these papers a different point of view from ours is taken: the SU(3)-structures are deduced from requiring that the intrinsic torsion of the $G_2$-structure vanishes on $G_2$-holonomy manifolds.  These $G_2$-structures are expressed in terms of SU(3) structures and the vanishing of the $G_2$ intrinsic torsion is then expressed in terms of the SU(3)-structures.  These papers focus on the interesting case where $(d\Omega)^{(2,2)}=0=F^{(2,2)}$.  Under these conditions the almost complex structure on $\cal M$ becomes integrable, thus $\cal M$ becomes a complex manifold and the metric $g$ (in our notation) is K{\" a}hler.  

We hope that the general constraints that we have written on a certain class of of $G_2$-holonomy manifolds (those related to wrapped D6-branes) lead to new explicit solutions to the constraints.  Even in the absence of such explicit solutions important results might still be obtainable.  For instance, we know \cite{amv} that wrapped D6-branes and their lift describe 4d N=1 gauge theory at two different energy scales - the ultraviolet perturbative and the infrared confined respectively.  It would be interesting to discover how the data that specify the manifolds relate to physics questions in gauge theories in more general scenarios beyond the conifold.  

\no
\\

\no
{\Large \bf Acknowledgements}\\
\no
AF would like to thank the physics departments at Harvard University and Stockholm University for hospitality during the course of this work and the Swedish Vetenskapsr{\aa}det (VR) for travel funds.  TZH would like to acknowledge funding from VR.  We would also like to thank Faheem Hussain and the National Center for Physics in Islamabad, Pakistan which hosted the 12th Regional Conference in Mathematical Physics for a stimulating conference and great working conditions.

%\newpage

}
\end{document}